\documentclass[useAMS,usenatbib,usegraphicx,amsmath,amssymb,graphicx,soul,color]{mn2e}
\newcommand{\aap}{A $\&$ A}
\newcommand{\apj}{ApJ}
\newcommand{\apjl}{ApJL}
\newcommand{\solphys}{SoPh}

\title[Homologous Active Region Jets]{Sunspot Waves and Triggering of Homologous Active Region Jets}
\author[R. Chandra, G. R. Gupta, Sargam Mulay, Durgesh Tripathi]{R. Chandra,$^{1}$\thanks{E-mail:
rchandra.ntl@gmail.com} G. R. Gupta$^{2}$, Sargam Mulay$^{2}$ and Durgesh Tripathi$^{2}$ \\
$^{1}$Department of Physics, DSB Campus, Kumaun University, Nainital 263002, India\\
$^{2}$Inter-University Centre for Astronomy and Astrophysics, Post Bag - 4, Ganeshkhind, Pune 411007, India}
\begin{document}

\date{Accepted ...... Received ........}

\pagerange{\pageref{firstpage}--\pageref{lastpage}} \pubyear{2002}

\maketitle

\label{firstpage}

\begin{abstract}
We present and discuss multi-wavelength observations of five homologous recurrent solar jets that occurred in active region NOAA 11133  on 11
December, 2010. These jets were well observed by the Solar Dynamic observatory (SDO) with high spatial and temporal resolution. The speed
of the jets ranged between 86 and 267~km~s$^{-1}$. A type III radio burst was observed in association with all the five jets. The investigation of
the over all evolution of magnetic field in the source regions suggested that the flux was continuously emerging on longer term. However, all the
jets but J5 were triggered during a local dip in the magnetic flux, suggesting the launch of the jets during localised submergence of magnetic 
flux. Additionally, using the PFSS modelling of the photospheric magnetic field, we found that all the jets were ejected in the direction of open field 
lines. We also traced sunspot oscillations from the sunspot interior to foot-point of jets and found presence of $\sim$ 3 minute oscillations 
in all the SDO/AIA passbands. The wavelet analysis revealed an increase in amplitude of the oscillations just before the trigger of the jets, that decreased after the jets were triggered. The observations of increased amplitude of the oscillation and its subsequent 
decrease provides evidence of wave-induced reconnection triggering the jets.
\end{abstract}

\begin{keywords}
Sun: jets -- Sun: magnetic field -- Sun: sunspots -- Sun: magnetic reconnection -- oscillations
\end{keywords}

\section{Introduction}

Solar jets are active phenomena in the solar atmosphere observed as collimated ejection of plasma. Jets were first discovered in the 
observations recorded by the Soft X-ray telescope \citep{Tsuneta91} onboard Yohkoh. Several studies have been performed on the 
estimation of different physical parameters of jets such as length, width, velocity and lifetime etc \citep[see e.g.,][]{Shimojo96,Shimojo00,Uddin12,Schmieder13}. 
The typical length of the jets ranges between 10$^4$ to 10$^5$ km, and the width varies from 10$^3$ to 10$^5$ km with a typical speed 
of about 200~km~s$^{-1}$. Often, these X-ray jets are observed along
with surges observed in H$\alpha$ filtergrams, which are essentially the cooler counterparts of the hot jets, observed at chromospheric
temperatures \citep{Schmieder95,Canfield96,Uddin04,Chandra06}. However, their different observational manifestations may not be
temporally and spatially aligned \citep{Chae99, Liu04}.

Observations across a range of wavelengths have provided evidences in support of magnetic reconnection models as the trigger of
solar jets (Heyvaerts et al. 1977; Yokoyama et al. 1995; Isobe $\&$ Tripathi 2006; Chifor et al. 2008a,b; Moreno-Insertis et al. 2008; 
Isobe et al. 2007; Pariat et al. 2009, 2010; Archontis $\&$ Hood 2012; Guo et al. 2013; Mulay et al. 2014).  Using the observations 
recorded by Hinode/EIS, Chifor et al. (2008a) studied the recurring jets of 15 \& 16 January, 2007 and measured temperature, 
Doppler shift, density, and filling factor in the jets. Their results of high density, small filling factor along with red and blue shifted 
emissions suggested the occurrence of multiple small-scale magnetic reconnection events leading to the production of observed 
jets.The models for Quasi-Separatrix Layers (QSLs) \citep{Mandrini96} and null point reconnection \citep{Torok09} have also been 
considered for the initiation of jets.

In a statistical study of active region jets, \cite{Shimojo96} found that the jets are most often observed towards the western side of
the leading spot. Later \citet{Shimojo00} and \citet{Pariat10} found multi-polar magnetic field configuration in the source regions, often having
anemone magnetic structure \citep{shibata1992,Shimojo96, Torok09}. An anemone magnetic structure is a type of active region
where one magnetic polarity is surrounded by opposite polarity regions. It was later suggested that observations of jets in active
regions with anemone magnetic topology provides strong indication of 3D null point coronal magnetic reconnection \citep{Torok09}.

Occasionally, jets show recurring and successive occurrence at the same location with very similar morphological features.
Such type of jets are called \textit{homologous jets}. By performing a  3D modeling of homologous jets, it was shown that the
homology of jets can be produced due to reconnection at 3D null points \citep{Pariat10}. According to \citet{Pariat10}, the homology
of the jets is the result of formation of 3D null-point topology system driven continuously by twisted motion. However, there are
studies, where the recurrent homologous jets occur without null point topology \citep{Mandrini96,Schmieder97,Guo13}. Therefore,
the cause of recurring homologous jets is a matter of strong debate.

While analysing quiet Sun transition region explosive events, \citet{2004A&A...419.1141N} found that in some cases, explosive
events within a burst were separated by 3-5 min, and suggested the role of the wave-induced reconnection in triggering the individual 
events. MHD simulations \citep[e.g.,][]{2006SoPh..238..313C, 2009ApJ...702....1H} have suggested that wave-induced reconnection 
is a viable mechanism for initiation of jets.  Recently \citet{innes11} reported the relationship between the triggering of a jet and sunspot 
oscillations. However, the role of the oscillations played in triggering of these jets is are still not clear.

In this paper, we present a multi-wavelength observations of homologous jets occurring in NOAA \textit{AR 11133} on 11 December,
2010. The active region was located at N15W23 on the solar disk on 11 December, 2010. In section~\ref{morpho}, we present the
observational characteristics of the jets followed by the structure and evolution of magnetic field in the source regions in Section~\ref{field}.
The associated type III radio bursts as observed by WIND/WAVES {\bf are} presented in section~\ref{dynamic}. We discuss
the dynamics of the associated sunspot and associated wave in section~\ref{oscillations}. Finally, we discuss and conclude our 
results in section~\ref{discussion}.

\begin{figure*}
\centering
\centerline{\includegraphics[width=1.1\textwidth,clip=]{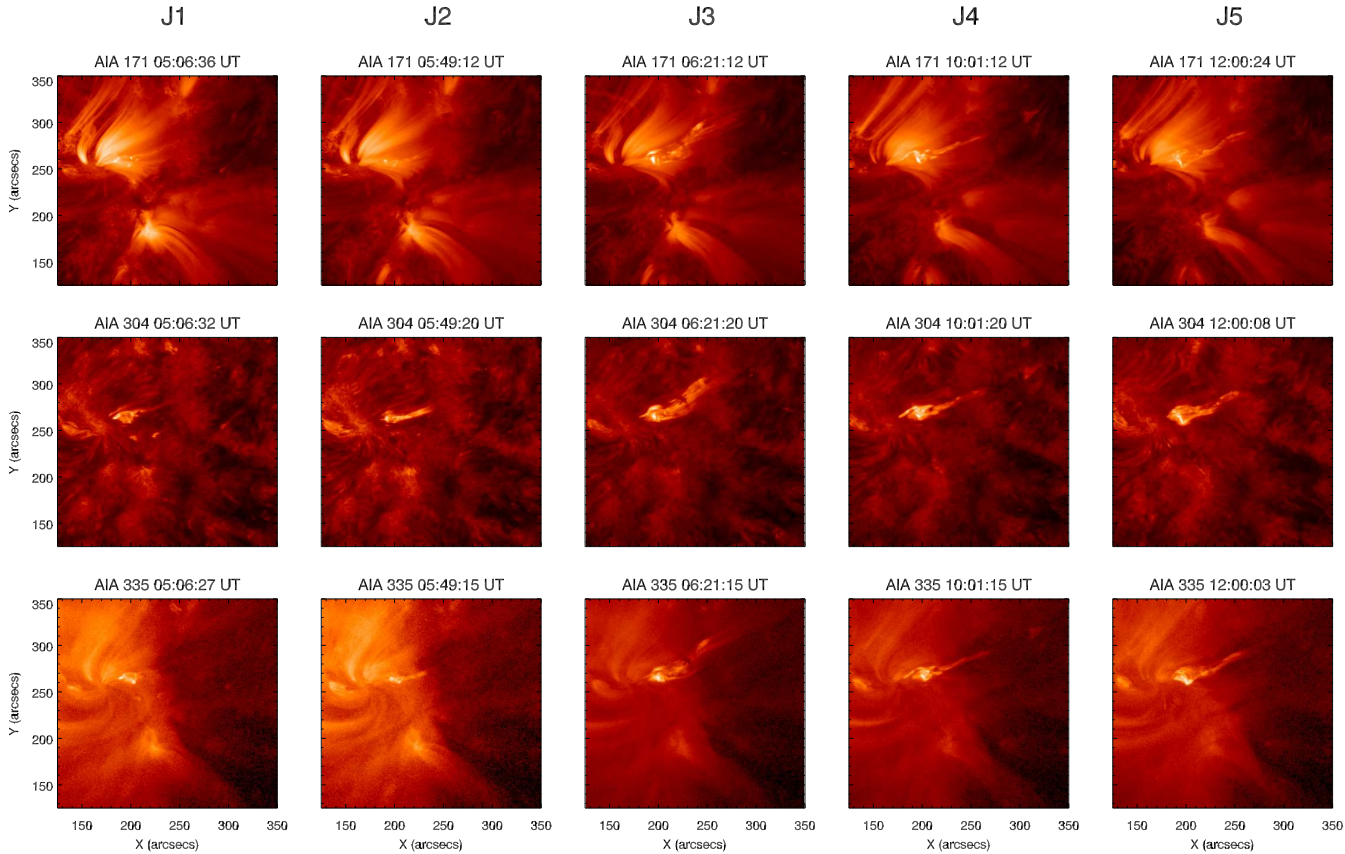}}
\caption{Peak phase of five jets i.e. J1, J2, J3, J4, and J5 observed by SDO/AIA at 171~{\AA}, 304~{\AA}, and 335~{\AA}.}
\label{evolution}
\end{figure*}
\section{Morphological, temporal and spatial evolution of jets} \label{morpho}
The Atmospheric Imaging Assembly \citep[AIA, ][]{lem12} onboard the Solar Dynamic Observatory \citep[SDO, ][]{Pesnell12} observes
the full disk of the Sun using seven different UV-EUV filters with an average cadence of 12~sec and pixel size of 0.6\arcsec. For the present
study, we have used the observations recorded by 171~{\AA}, 304~{\AA} and 335~{\AA} filters. Figure~\ref{evolution} displays the images of
the five jets observed at 171~{\AA} (first row), 304~{\AA} (second row), and 335~{\AA} (third row). The five jets are labelled in the figure as
J1, J2, J3, J4 and J5. The AIA observations are processed using standard processing software as provided in the solar software distribution.
The AIA observations show that all the observed five jets are homologous. They originate from the same location and ejected in the
same north-west direction except the jet J3, which was directed a bit northward than the others.

\begin{figure*}
\vspace*{-0.5cm}
\centerline{\includegraphics[width=0.90\textwidth,angle=-90,clip=]{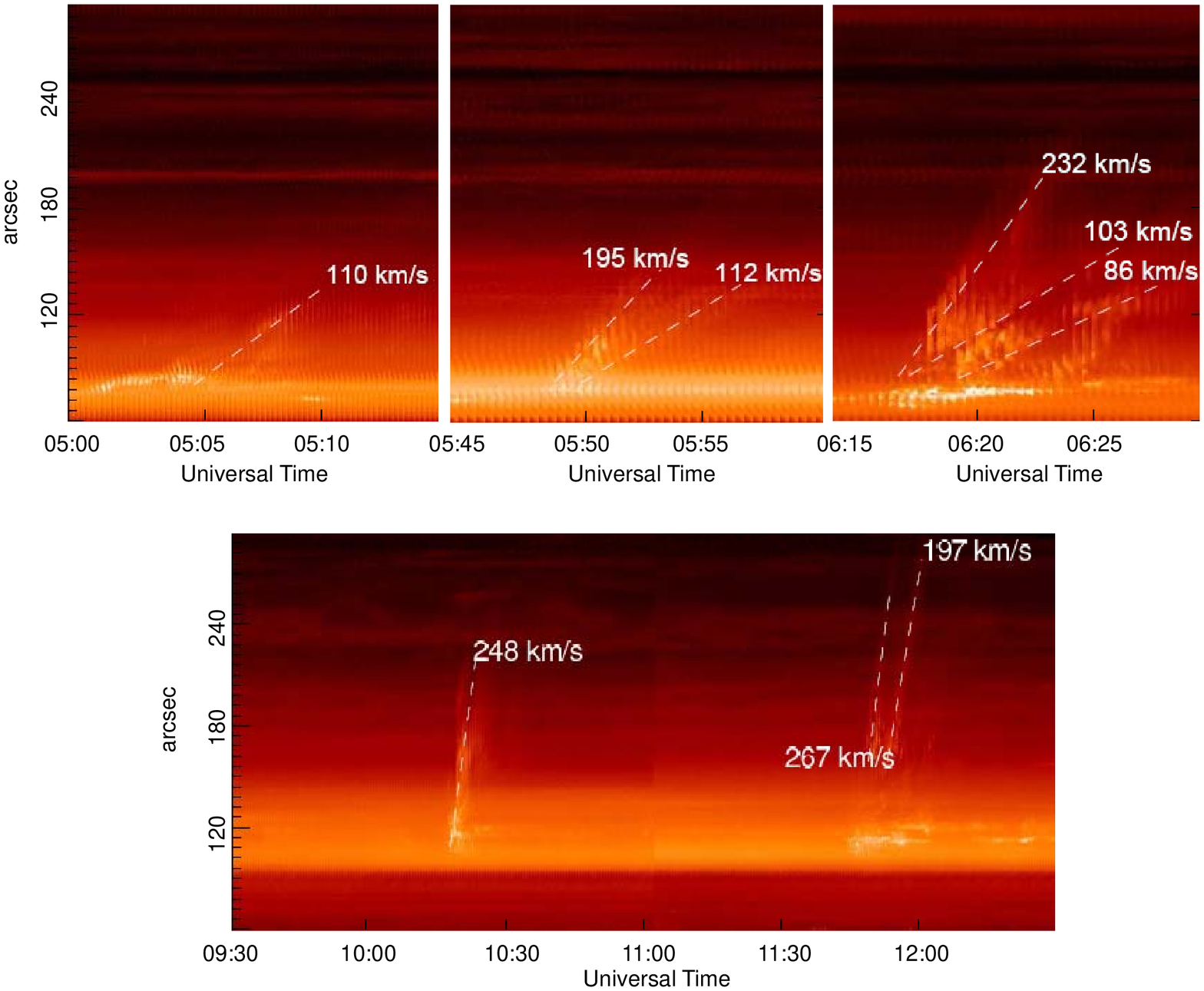}}
\caption{Time-Distance images through the jet 1 (top, left), jet 2 (top, middle), jet3 (top, right) and jet 4, jet 5 (bottom) along with linear fit. The slits are taken along the direction of jet ejection.}
\label{timeslice}
\end{figure*}

In order to estimate the speeds of the jets, we have performed the study of kinematics of the jets by using the time-slice techniques on the
171~{\AA} data. As mentioned earlier, since all the five jets are not exactly in the same direction, it is not feasible to employ a slit in one fixed
direction, which can study the time-slice analysis for all given jets. Therefore, it is mandatory to employ different slits for different jets. For all
the jets, the slit is aligned along the direction of plasma flow and centred on the jet. Figure~\ref{timeslice} show the height--time diagram for 
all five jet events as obtained by time-slice analysis. For the first jet `J1' (shown in the top left panel), the speed is 110~km~s$^{-1}$ and the
projected height of the jet is 95~Mm. In the case of `J2' jet, the computed speeds are 195 and 112 km~s$^{-1}$ and the projected height is 
96~Mm, which is close to the projected height of jet `J1'. Jet `J3' shows the time-slice pattern of different speeds ranging from 86 to 
232~km~s$^{-1}$. The maximum height attained by the jet `J3' is 143 Mm. The projected speeds for jets `J4' and `J5' are 248 and 
267~km~s$^{-1}$ and the maximum heights ares 165 and 204 Mm respectively. Our above computed speeds are comparable to earlier observed 
speeds \citep[see e.g.,][]{Shimojo96,Cirtain07,Patsourakos08,Schmieder13}.

\begin{figure*}
\vspace*{-0.5cm}
\centerline{\includegraphics[width=0.60\textwidth,clip=]{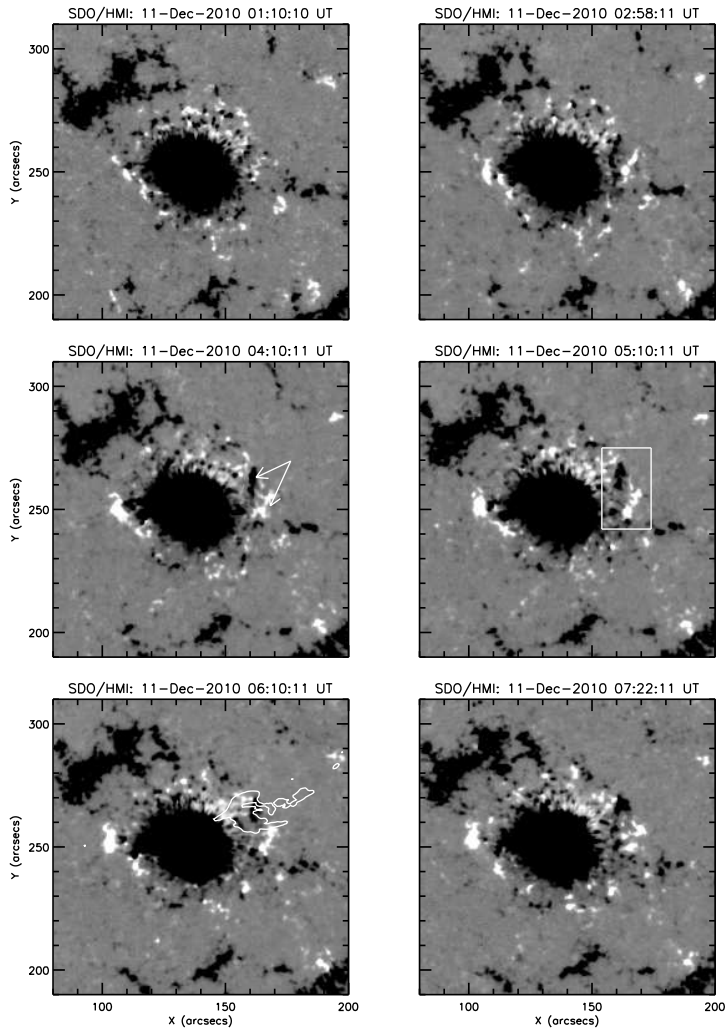}}
\caption{Evolution of HMI magnetic field on 11 December, 2010. Shown by arrows are the positive as well as negative flux emergence observed
before the jets. The white contours overlaid at 06:10:11 UT represent the location of J3 jet.}
\label{magnetic}
\end{figure*}

\begin{figure*}
\centerline{\includegraphics[width=0.50\textwidth,clip=]{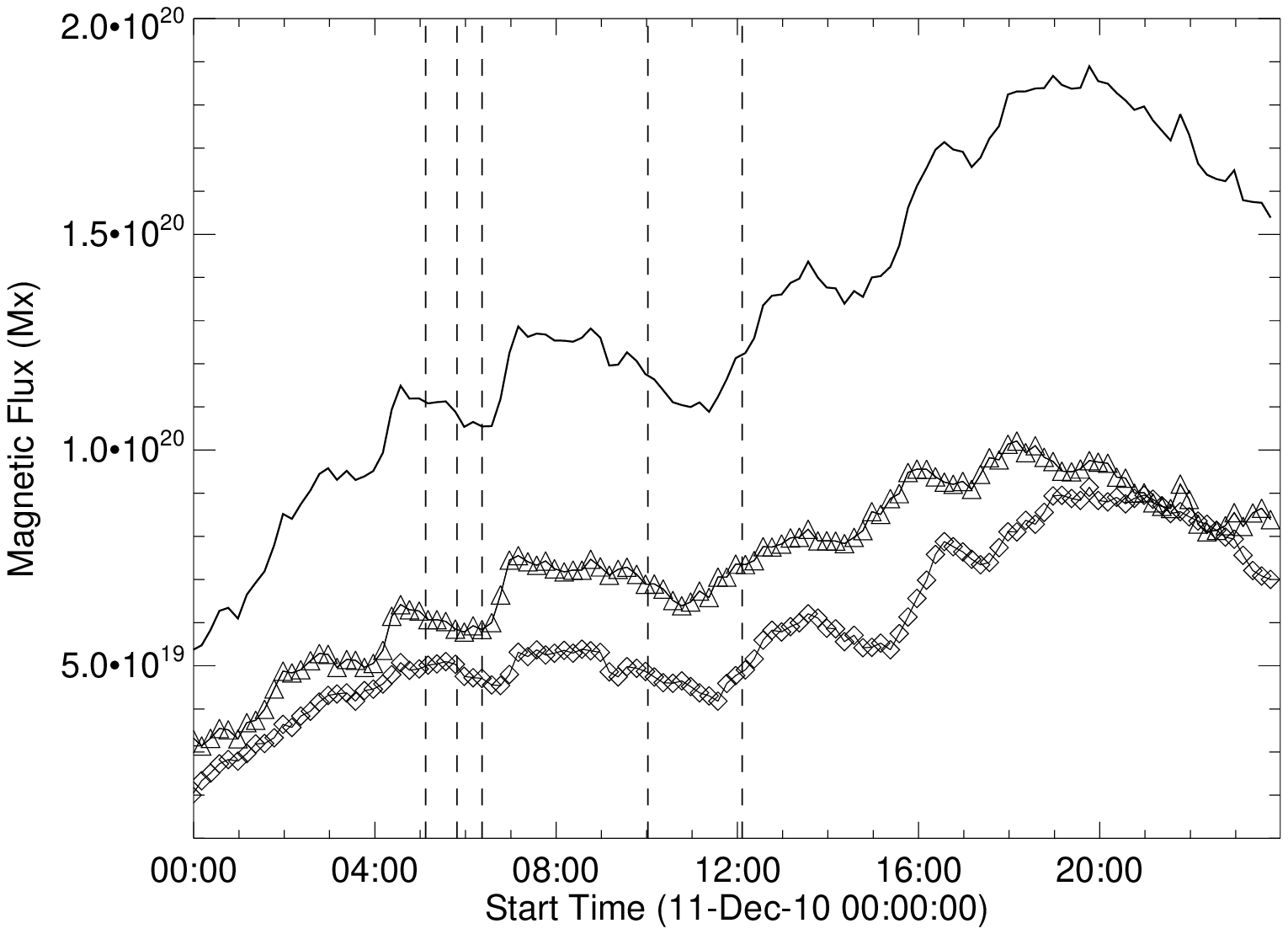}}
\caption{Evolution of HMI magnetic flux, computed in the box drawn in figure \ref{magnetic} (middle, right).
Diamond, triangle and solid line indicate positive, unsigned negative and total unsigned magnetic flux
respectively. Vertical dashed lines indicate the time of jets J1 to J5 respectively.}
\label{flux}
\end{figure*}
\section{Evolution of Magnetic field} \label{field}
In order to understand the magnetic cause of the jets, we studied the photospheric magnetic field configuration in their source regions using the 
observations recorded by the Helioseismic and Magnetic Imager (HMI) onboard SDO. Here we only study the line of sight component of the 
magnetic field, shown in Figure~\ref{magnetic}. The active region in which the jets were observed appeared on the east limb on 04 December, 
2010 and went over the west limb on 16 December, 2010. The active region is comprised of a main negative polarity surrounded by 
numerous small positive polarity regions, similar to the moat regions of sunspots during their decay phase. (Brooks et al. 2007). A negative part of 
the sunspot is detached from the main polarity and increased in size and flux. Then it progressively cancelled with positive polarities. These observations are similar to those reported by  \citet{Pariat04} from Flare Genesis Experiment.

The time evolution of the HMI magnetograms reveals emergence of both positive as well as negative fluxes co-spatial to the location of the jets. The 
emergence of magnetic flux at the jet location is shown by white arrow in Figure \ref{magnetic} at 04:10:11~UT on 11 December, 2010. The foot-points 
of all the five jets were located at the western edge of the active region. As an example, the location of one of the jets (J3) is overlaid on magnetic field 
image at 06:10:11~UT as contour in the bottom left panel of Figure \ref{magnetic}.

In order to quantify the magnetic flux at the jet basis, we have computed the flux in a box, shown in the middle right panel of Figure~\ref{magnetic}. 
The evolution of the magnetic flux is shown in Figure~\ref{flux}. The plot shows overall trend of flux emergence upto 20:00 UT on 11 December, 2010 
including the time interval of the jets. However, a closer look at the flux evolution and the times of jets appearance, we find that the jets `J1', `J2', `J3' and `J4'
were launched during the decrease in both positive and negative magnetic fluxes, where as jet `J5', was launched during the enhancement of magnetic flux.

\begin{figure*}
\centerline{\includegraphics[width=0.60\textwidth,clip=]{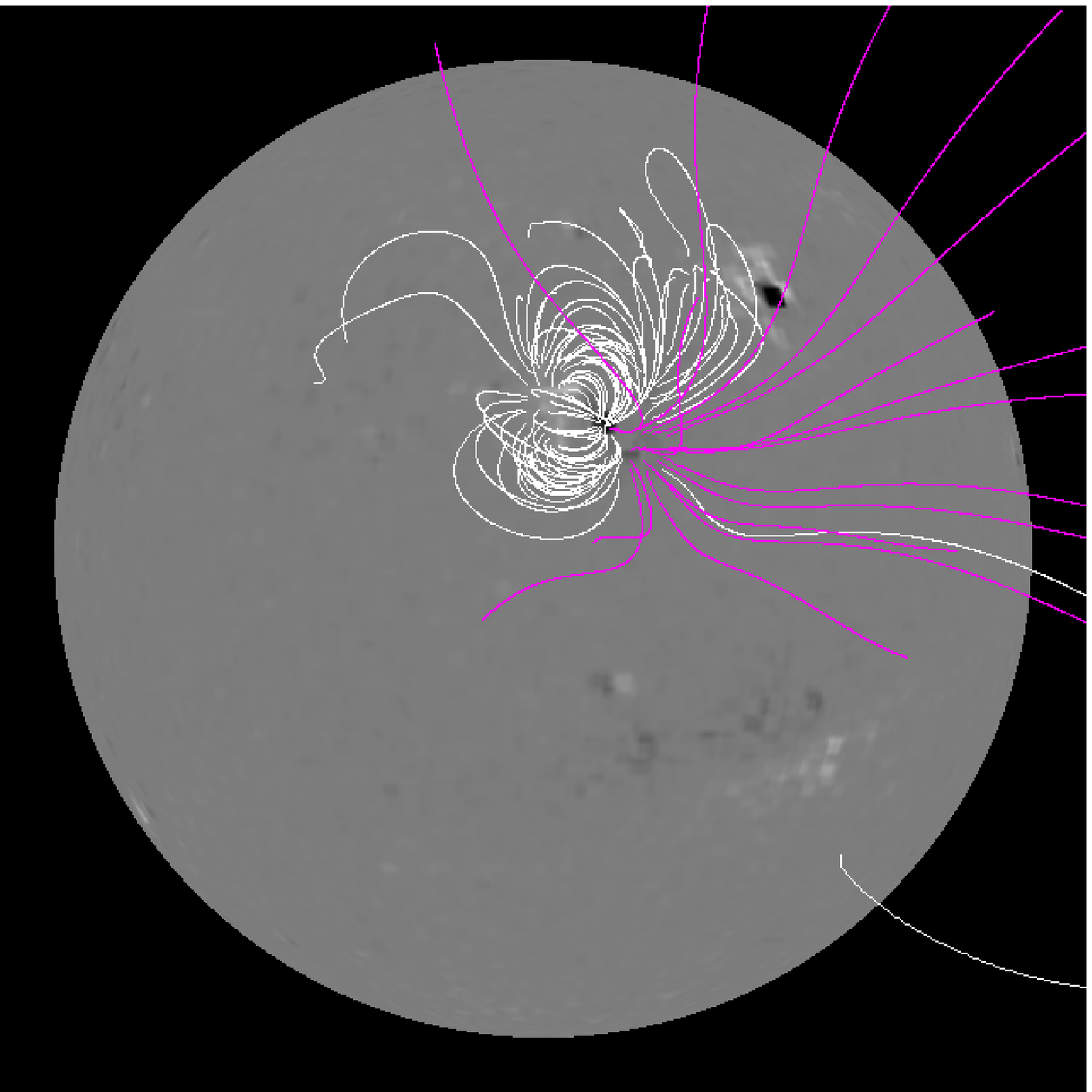}}
\caption{PFSS extrapolation of active region at 00:04:00 UT on 11 December, 2010. The white and pink field 
lines represent the closed and open field lines respectively.}
\label{pfss}
\end{figure*}

We have also investigated the overall magnetic topology of the active region using the Potential Field Source Surface (PFSS) extrapolation of
photospheric magnetic field observed by HMI at 00:04:00~UT using the software provided in SSW. The PFSS extrapolation is presented in
figure \ref{pfss} and reveals the presence of close as well as open field lines in the active region. A close comparison of the EUV observations with the
extrapolated field reveals that the jets are ejected in the direction of open field lines.

\section{Associated type III radio bursts} \label{dynamic}
\begin{figure*}
\centerline{\includegraphics[width=0.80\textwidth,clip=]{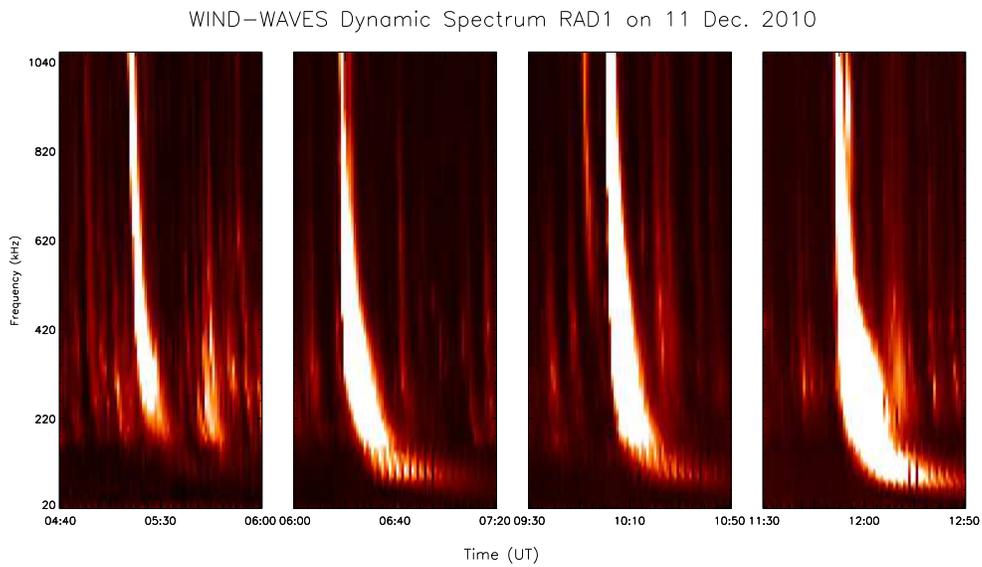}}
\caption{Dynamic spectrum of the observed jets as recorded by {\it WIND/WAVES} on 11 December, 2010. The first panel corresponds to
jets J1 and J2. The 2nd, 3rd and 4th panels correspond to the jets J3, J4 and J5.}
\label{radio}
\end{figure*}

Observations of type III radio bursts represent the escaping non-thermal electrons along the open field lines and provides evidence of magnetic
reconnection. We compared our observation with those recorded with {\it WIND/WAVES} \citep{Bougeret95}. The dynamic spectrum from
20 to 1040 kHz as observed by WIND/WAVE using RAD 1 instrument is displayed in Figure~\ref{radio}. The 1st panel corresponds to J1
and J2, while the 2nd, 3rd and 4th panels correspond to J3, J4 and J5. The comparison between the jet start time and detection
time of type III burst reveals a strong correlation. We find that all the jets are associated with type III radio burst.

\section{Sunspot oscillations}\label{oscillations}
\begin{figure*}
\hbox{\centerline{\includegraphics[width=0.9\textwidth,clip=]{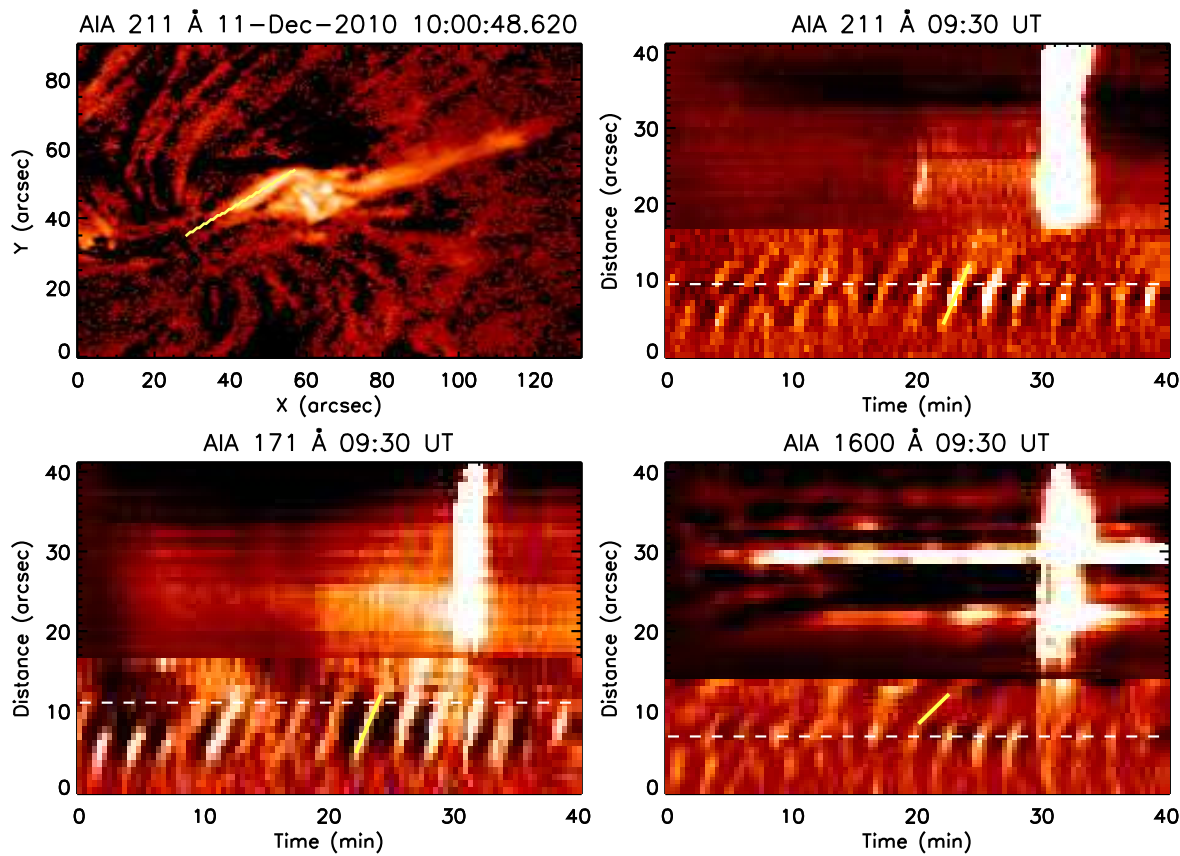}}}
\caption{A jet event J4 observed in AIA 211 \AA\ passband (top left). Artificial slit location is also
overplotted for which time-distance maps were obtained for further analysis. In rest of the panels,
time-distance maps are plotted for the slit location as obtained in AIA 211 \AA\ (top right), 171 \AA\
(bottom left), and 1600 \AA\ (bottom right) passbands. Two parts of time-distance maps obtained from
artificial slit (a part covering sunspot and another covering jet beam) are processed differently so
as to bring out the propagating features nicely. Sunspot waves can be seen propagating towards
the jet foot point with speed 46 km s$^{-1}$ , 44 km s$^{-1}$ , and 18 km s$^{-1}$  in AIA 211 \AA , 171 \AA , and 1600\AA\ 
passbands respectively with period about 3 minutes. Time series obtained at over-plotted dashed lines
are used for wavelet analysis (see Fig.~\ref{fig:wvlet}).}
\label{fig:loc_xt}\end{figure*}

In order to understand the relationship between sunspot waves and jets, we studied sunspot oscillations originating from the interior and
propagating towards the location of the jet. We have performed this analysis for the all five jet events. However, we present the results for
the jet J4 as this is the strongest among all (see Figure~\ref{fig:loc_xt}).

We drew artificial slit originating from sunspot interior and covering the jet beam and obtained time-distance maps along the slit in AIA 
211~{\AA}, 171~{\AA}, and 1600~{\AA} passbands (see Figure~\ref{fig:loc_xt}). Two parts of time-distance maps obtained from 
artificial slit (a part covering sunspot and another covering jet beam) are processed differently so as to bring out the propagating features 
unambiguously. The time-distance maps obtained in different passbands reveal the presence of quasi-periodic propagating disturbances 
originating from sunspot and propagating towards the jet foot-point. The propagation speeds of the waves are 46 km~s$^{-1}$, 44 km~s$^{-1}$ 
and 18 km~s$^{-1}$ observed in AIA 211~\AA\, 171~\AA\, and 1600~\AA\  respectively as obtained from the slope of bright ridges.

\begin{figure*}
\hbox{\hspace*{-4cm}\centerline{\includegraphics[width=0.35\textwidth,angle=90,clip=]{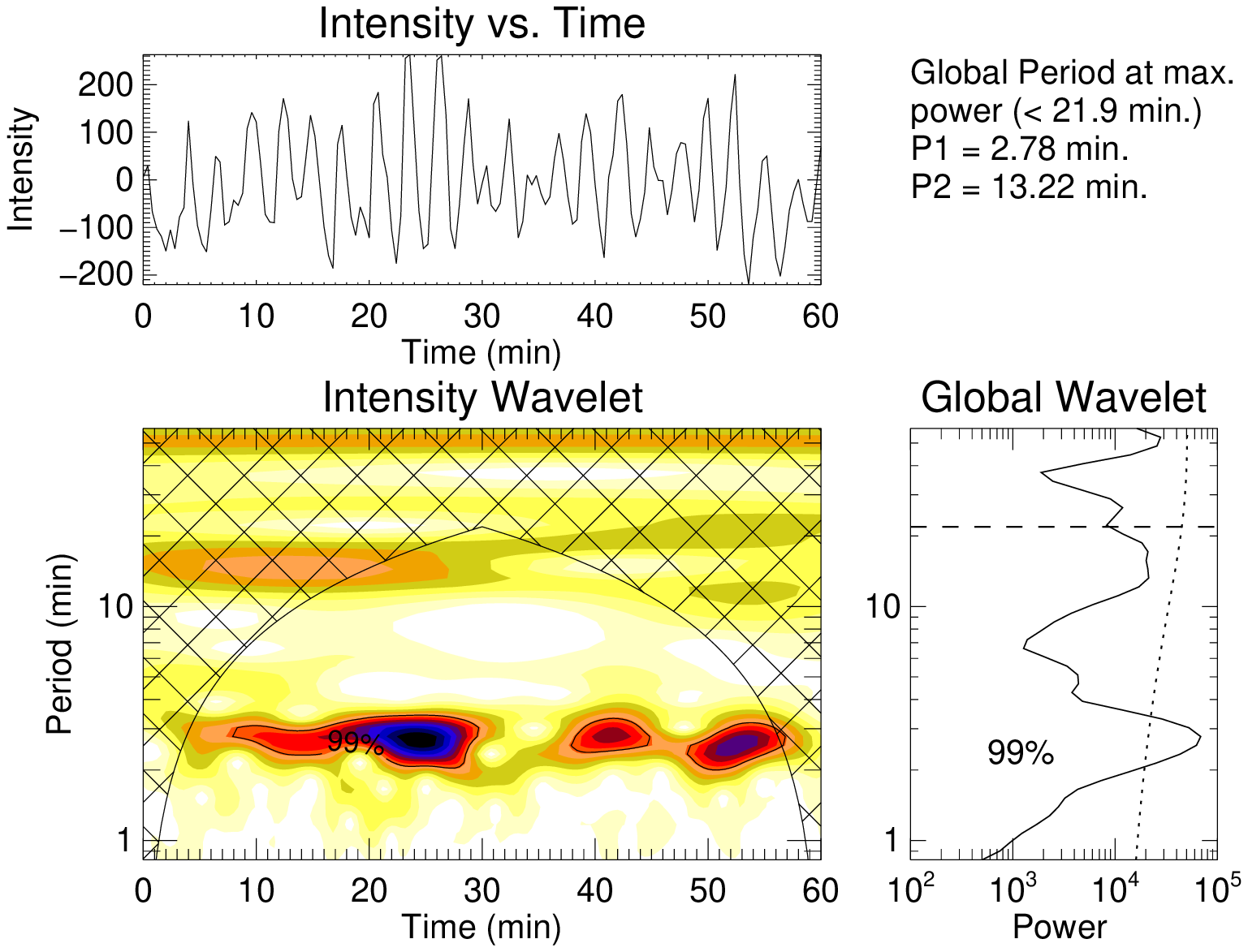}}
\hspace*{-9cm}\centerline{\includegraphics[width=0.35\textwidth,angle=90,clip=]{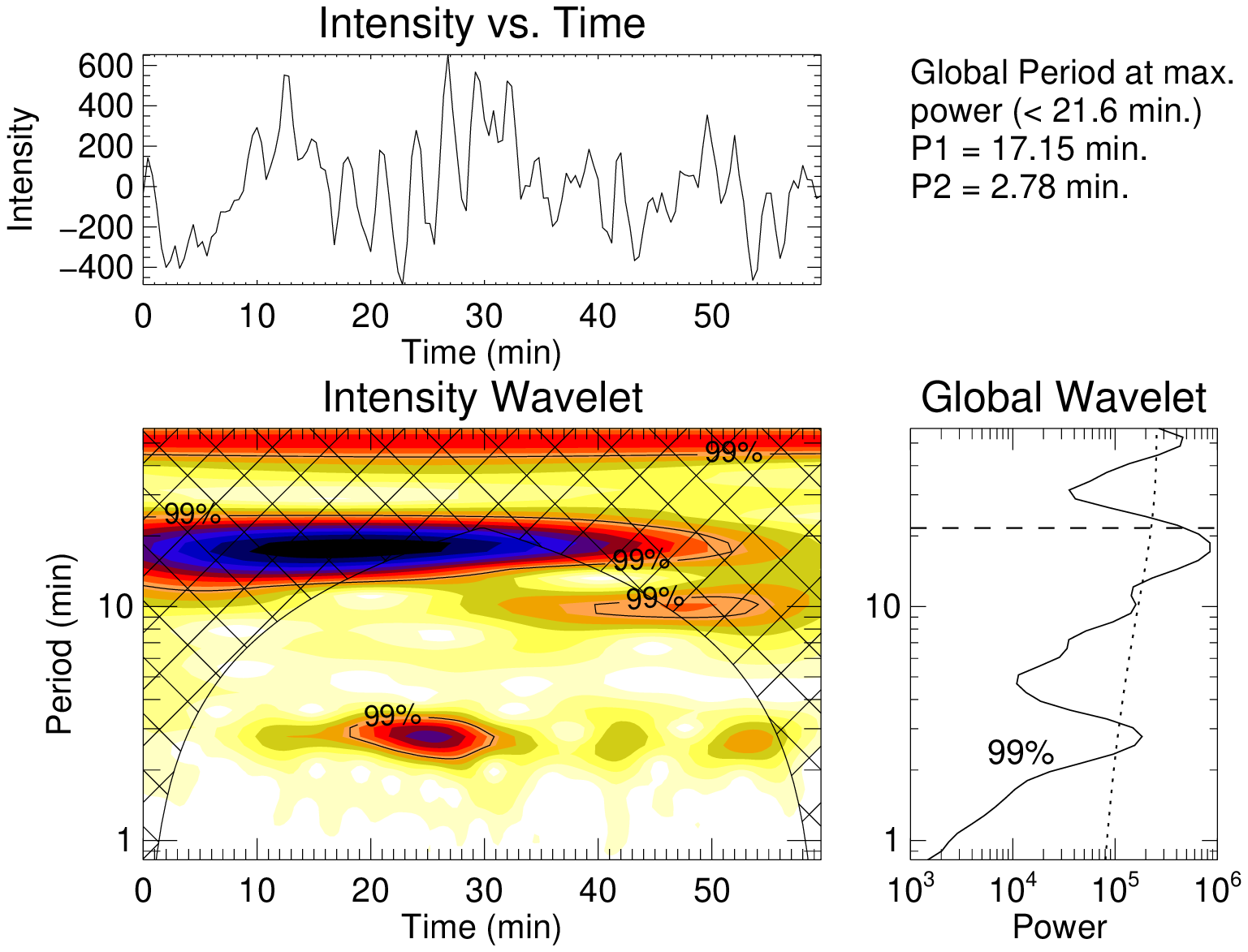}}}
\caption{The wavelet analysis results for the locations marked with dashed lines in time-distance maps (See Fig.~\ref{fig:loc_xt}) for
211~{\AA} (left) and 171~{\AA} (right) passbands. In each sets, the top panels show the variation of trend subtracted (50-point running average)
intensity with time. The bottom-left panels show the colour-inverted wavelet power spectrum with 99 \% confidence-level contours,
while the bottom-right panels show the global wavelet power spectrum (wavelet power spectrum averaged over time) with 99 \% global
confidence level drawn. The periods P1 and P2 at the locations of the first two maxima in the global wavelet spectrum are shown above
the global wavelet spectrum.}
\label{fig:wvlet}
\end{figure*}

In order to further understand the characteristics of the sunspot oscillations, we have also performed wavelet analysis \citep{1998BAMS...79...61T} with the 
time series obtained at the locations marked with dashed lines on the time-distance maps. We have performed the analysis
with all the three passbands, but show only 211~{\AA} (left panel) and 171~{\AA} (right panel) passbands. Figure~\ref{fig:wvlet} displays the obtained
results. The top panels show the variation of intensity with time. Note that a 50 point running average has been subtracted from the original curve to
bring out the oscillations more clearly. The bottom left panels are wavelet power spectrum (colour inverted) with 99\% confidence levels
and bottom right panels are global wavelet power spectrum with 99\% global confidence. As expected, the wavelet power spectrum clearly
reveals the presence of $\approx$ 3 min oscillations in all the passbands.

\begin{figure*}
\hbox{\centerline{\includegraphics[width=0.80\textwidth,clip=]{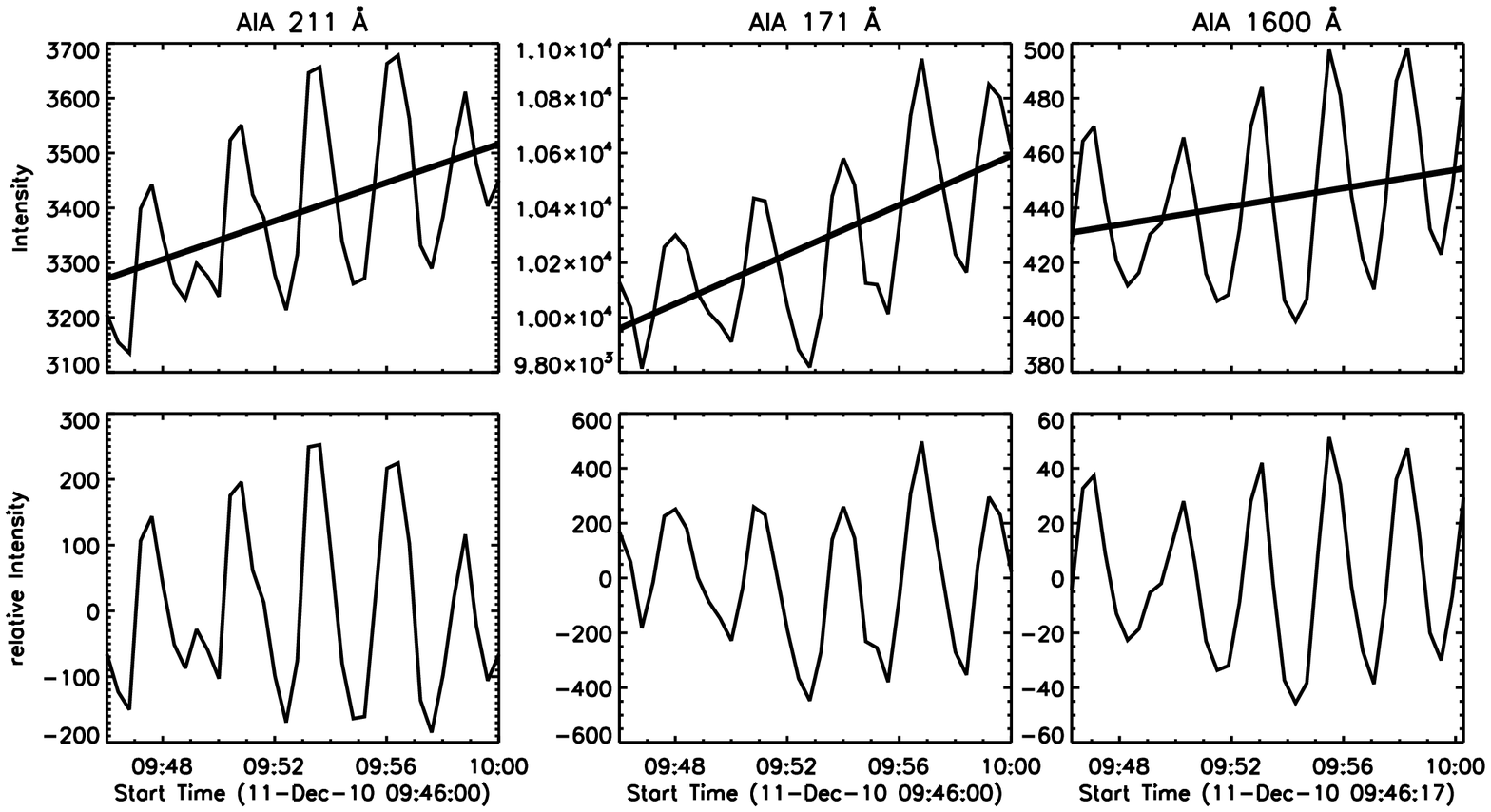}}}
\caption{Intensity variation just before jet J4 event at locations marked with dashed lines in time-distance maps (See Fig.~\ref{fig:loc_xt})
observed in AIA 211, 171, and 1600 \AA\  passbands (Top panels). Overplotted continuous lines  are linear fit applied to the intensity variation.
Bottom panels: Relative intensity variation with time obtained after subtracting the linear trend (fit) from the original intensity for the different AIA passbands as labelled. Relative intensity variation with time in different passbands clearly show the increase in amplitude of wave oscillations prior to the jet event J4. }
\label{fig:lc_amp}
\end{figure*}

Moreover, a very interesting and unique feature is observed in this data set, i.e. an increase in the amplitude of the oscillation in all the passbands 
right before the jets were triggered. This can be visualised in the wavelet time-series plots where an increase in oscillatory power is observed from 
10 minutes onwards (see top left panel in Fig.~\ref{fig:wvlet}). To understand this phenomena in greater details, we chose small interval of the time 
series  and plotted in Figure~\ref{fig:lc_amp}. A linear fit to obtain the trend in the intensity variation for this small time interval is applied to the intensity 
variation and over plotted. Relative intensity variations with time were obtained after subtracting the linear trend (fit) from the original intensity for the 
different AIA passbands, which clearly shows the increase in the amplitude of oscillations prior to the jet event J4. A similar characteristic was observed
for all the other jets.

\section{Summary and Discussion} \label{discussion}
In this study, we have investigated multi-wavelength observations of five recurrent homologous jets. These jets occurred in the active
region NOAA AR 11133. The main results of our study are summarized as follows:

\begin{itemize}

\item The observed jets occurred in the western periphery of the active region similar to the observations by \cite{Shimojo96} and are homologous.
All the five jets are observed in all the channels of AIA suggesting multi-temperature structure of the jets. The time distance analysis provided the
speed of the jets which ranges between 86 and 267~km s$^{-1}$. These speeds are similar to those reported earlier.
\\
\item The magnetic field data show continuous evolution with emerging and submerging flux regions around the sunspot. Both positive and 
negative flux regions emerged at the location of jets just before the appearance of the first jet. There are various scattered small scale
field regions (both positive and negative) observed around the strong negative polarity sunspot. On a close comparison of jet timings with evolution 
of magnetic flux, we found that the jets `J1', `J2', `J3' and `J4' were associated with flux cancellation, while the jet `J5' was observed during the 
period of emerging magnetic flux. Additionally, the direction of ejection of jets are closely related to the open field lines as was revealed by the PFSS 
extrapolation of the photospheric magnetic field.
\\
\item All the jets were associated with type III radio bursts. It is believed that the non-thermal particles are accelerated at reconnection site in the
current sheet, which forms between closed and open magnetic field lines \citep{shibata1992, innes11} and excite the plasma oscillations
(Langmuir waves). These plasma oscillations are further converted to escaping electromagnetic radiation by non-linear wave-wave interactions and
produce its signatures in the radio dynamic spectrum as a 'Type III radio burst'. Therefore, the strong association between jets and
type III radio bursts provides further evidence in support of magnetic reconnection models in the production of jets.
\\
\item The analysis of the sunspot waves revealed a 3 minute oscillatory pattern. The analysis further revealed that an increase in the amplitude of
the oscillations prior to the triggering of the jet. This was observed for all the events and all the three AIA channels studied here. The oscillatory amplitudes
decreased after the launch of the jets.
\end{itemize}

The emergence and submergence of the magnetic flux regions in close temporal and spatial association with the jets along with the observations
of type III burst provides strong evidence supporting the reconnection model for the triggering of the jets. The observations of the increasing
and decreasing oscillatory power of the sunspot waves before and after the triggering of the jets respectively, provide evidence of wave induced magnetic
reconnection \citep[e.g.,][]{2006SoPh..238..313C, 2009ApJ...702....1H}. Increase of oscillatory power is related to increase of wave amplitude. The
increasing and decreasing of wave amplitude can be interpreted as expansion and contraction of flux tube. The increase in the amplitude will therefore
bring the neighbouring magnetic field lines closer, which may induce reconnection leading to the formation of jets. Similar wave induced
reconnection phenomena have also been speculated by \citet{2004A&A...419.1141N} while studying the quiet Sun transition region explosive events.
\citet{2009A&A...505..791S} also found a gradual increase in the amplitude of the 3-min oscillatory train in the sunspot before the triggering
of flares. It was proposed that flaring energy release was triggered by 3-min slow magneto-acoustic wave leakage from the sunspots,
similar to our observations of jet triggering. Thus, study of propagation of sunspot waves provide an important clue in our understanding
of triggering of jet events at the boundary of active regions.

\section*{Acknowledgments}

We thank the anonymous referee for valuable and helpful comments and suggestions.
The authors thanks the open data policy of SDO, and WIND/WAVES teams. RC acknowledges support from ISRO/RESPOND
project no. ISRO/RES/2/379/12-13. GRG acknowledges Department of Science and Technology (DST), India for INSPIRE Faculty Fellowship.
SM and DT acknowledge support from DST under the Fast Track Scheme (SERB/F/3369/2012/2013).


\end{document}